\documentclass[a4paper]{article}
\pdfoutput=1
\usepackage{icrc2013}
\usepackage[english]{babel}

\title{GeoMag and HelMod webmodels version for magnetosphere and heliosphere transport of cosmic rays}

\shorttitle{GeoMag \& HelMod webmodels}

\authors{
P. Bobik$^{2}$,
M. J. Boschini$^{1,3}$,
C. Consolandi$^{1}$,
S. Della Torre$^{1}$,
M. Gervasi$^{1,4}$,
D. Grandi$^{1}$,
K. Kudela$^{2}$,
G. La Vacca$^{1,4}$,
S. Pensotti$^{1,4}$,
M. Putis$^{2}$,
P.G. Rancoita$^{1}$,
D. Rozza$^{1,5,6}$ and
M. Tacconi$^{1,4}$
}

\afiliations{
$^1$ INFN Milano Bicocca - Italy \\
$^2$ Institute of experimental Physics, Kosice - Slovakia\\
$^3$ CINECA \\
$^4$ University of Milano Bicocca \\
$^5$ University of Insubria \\
$^6$ CERN \\
}

\email{Stefano.DellaTorre@mib.infn.it}

\abstract{
We implemented a website to deal with main effects on Cosmic Ray access to the Earth, i.e. the Solar Modulation (\cite{bib:helmod}) 
and the Geomagnetic Field effect (\cite{bib:geomag}). In helmod.org the end user can easily access a 
web interface to results catalog of the HelMod Monte Carlo Code. This Model uses a Monte Carlo Approach 
to solves the Parker Transport Equation, obtaining a modulated proton flux for a period (monthly average) 
between January 1990 and december 2007. 

geomagsphere.org is instead based on GeoMag Backtracing Code, that solves the Lorentz equation with a Runge-Kutta method of 6th order, 
and, reversing charge sign and velocity, reconstruct particle trajectories in the Earth Magnetosphere back in time. We use last models 
of internal (IGRF-11) and external (Tsyganenko 1996 -T96- and 2005 -T05-) field components valid up to 2015. 

Particles are backtraced to the outer 
(magnetopause) or inner boundary to separate Primary (allowed trajectory) from Secondary (forbidden) Cosmic Rays. 
This code has been used both for reproducing known effects as East-West effect and rigidity cutoff calculations. 
In geomagsphere.org the user can choose the external field model from Tsyganenko (T96 or T05) and obtain for a fixed position and date 
from 1$^{st}$ Jan. 1968 (T96) and 1$^{st}$ Jan. 1995 (T05) respectively till 31$^{st}$  Dec 2012, the vertical rigidity cutoff estimation obtained with 
the backtracing technique with a rigidity step of 0.1 GV. For a more precise calculation (0.01 GV), requiring more CPU time, results 
are sent to the user by email (mail model). 
}


\begin{document}
\maketitle

\section{Introduction}
The local interstellar spectrum behind the heliosphere borders is isotropic and constant in time. Cosmic rays detected on the Earth surface, 
in the atmosphere or at Earth orbits are modulated by the heliospheric magnetic field and the geomagnetic field. Particles in the heliospheric magnetic 
field are scattered on irregularities and adiabatically lose energy. Particles inside the magnetosphere are redistributed 
in direction by the Lorentz force but do not lose energy. Thus two models are needed to describe ``the journey''  of a 
particle from interstellar space to a detector taking data inside Earth magentosphere. 

The first one is a model of particle propagation from the heliopause through the heliosphere to a distance of 1 AU
from the Sun. The second one evaluates the propagation from the magnetopause to the detecting position inside the magnetosphere. 
We implemented a couple of websites to offer to the user access to developed models that deal with these 
main effects on Cosmic Ray access to the Earth, i.e. the Solar Modulation (\cite{bib:helmod}) and the Geomagnetic 
Field effect (\cite{bib:geomag}).

\section{Geomagsphere.org }

Geomagsphere.org is based on the GeoMag Backtracing Code \cite{bib:ams_mib}, that solves the Lorentz equation with a 
Runge-Kutta method of 6$^{th}$ order, 
by reversing charge sign and velocity and reconstruct particle trajectories in the Earth Magnetosphere back in time. 

We used the last models of internal (IGRF-11 \cite{bib:IGRF}) and external (Tsyganenko 1996 -T96- and 2005 -T05- \cite{bib:T96,bib:T05}) 
geomagnetic field components valid up to 2015. 

In geomagsphere.org the user can choose the external field model from Tsyganenko (T96 or T05) and obtain for a fixed 
position and date the vertical rigidity cutoff estimation calculated with the backtracing technique with a rigidity step of 0.1 GV. 
For a more precise calculation (so a smaller step of 0.01 GV) that will require more CPU time, results are sent to the user by email 
(mail model). 

Web models GeoMag96 and GeoMag05 calculate trajectories for vertically incoming particles 
at selected position and time in the Earth's magnetosphere. Position is defined by a geographical latitude and longitude and a radius 
(from the Earth surface till 10 Earth radii). After setting position and time by the user, the system automatically find corresponding input 
parameters (from OMNIWeb database \cite{bib:omni}) needed for simulation. For model GeoMag96 those are the Dst index, dynamic pressure of solar 
wind Pdyn and Y and Z component of interplanetary magnetic field. For model GeoMag05 Dst index, Pdyn, B$_{Y}$ and B$_{Z}$ 
are completed with W1, W2, W3, W4, W5 and W6 parameters \cite{bib:05REP} 
describing prehistory of geomagnetic field in the last 24 hours before the user selected simulation moment/time.

Model GeoMag96 with Tsyganenko 96 external field can be used in a period ranging from 1$^{st}$ january 1968 till 31$^{st}$ december 2012. 
Model GeoMag05 with Tsyganenko 05 external field is valid for from 1$^{st}$ january 1995 till 31$^{st}$ december 2012.

\begin{figure}[ht!]
  \centering
  \includegraphics[width=0.45\textwidth]{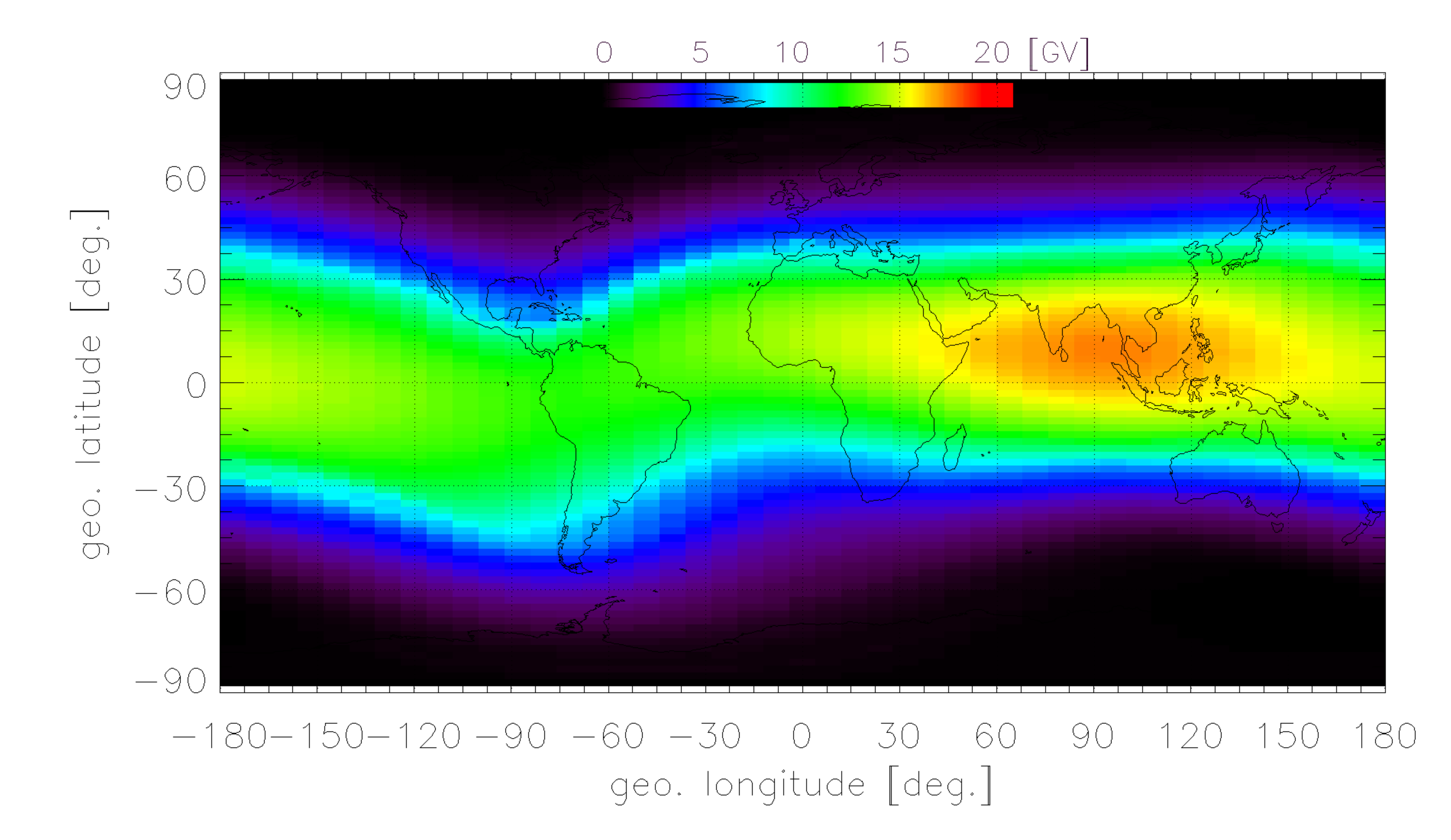}
  \caption{Plot of effective vertical cut-off rigidities at 20. march 2004 12:00 UTC.}
  \label{simp_fig1}
\end{figure}

Simulation run by the user from web interface provides a set of output values for allowed trajectories \cite{bib:sim}. 
For every trajectory these are rigidity 
in GV, proton velocity, position of trajectory end point at magnetopause, coordinates of asymptotic direction of particle, 
time spent in magnetosphere and length of trajectory inside magnetosphere. Also low, upper and effective cut-off rigidity is evaluated.
Example of model results is presented at Figure \ref{simp_fig1}, where it is shown the effective vertical cut-off rigidity map on Earth surface for 
20$^{th}$ march 2004 12:00 UTC evaluated in model with Tsyganenko 96 external geomagnetic field.  Map contains cut-off rigidities for 
4050 positions (grid with 2 deg. in geographical latitude and 8 deg. in geographical longitudes). Table with data used to create map of cutoff 
rigidities can be downloaded from geomagsphere.org.

\begin{figure}[ht!]
  \centering
  \includegraphics[width=0.45\textwidth]{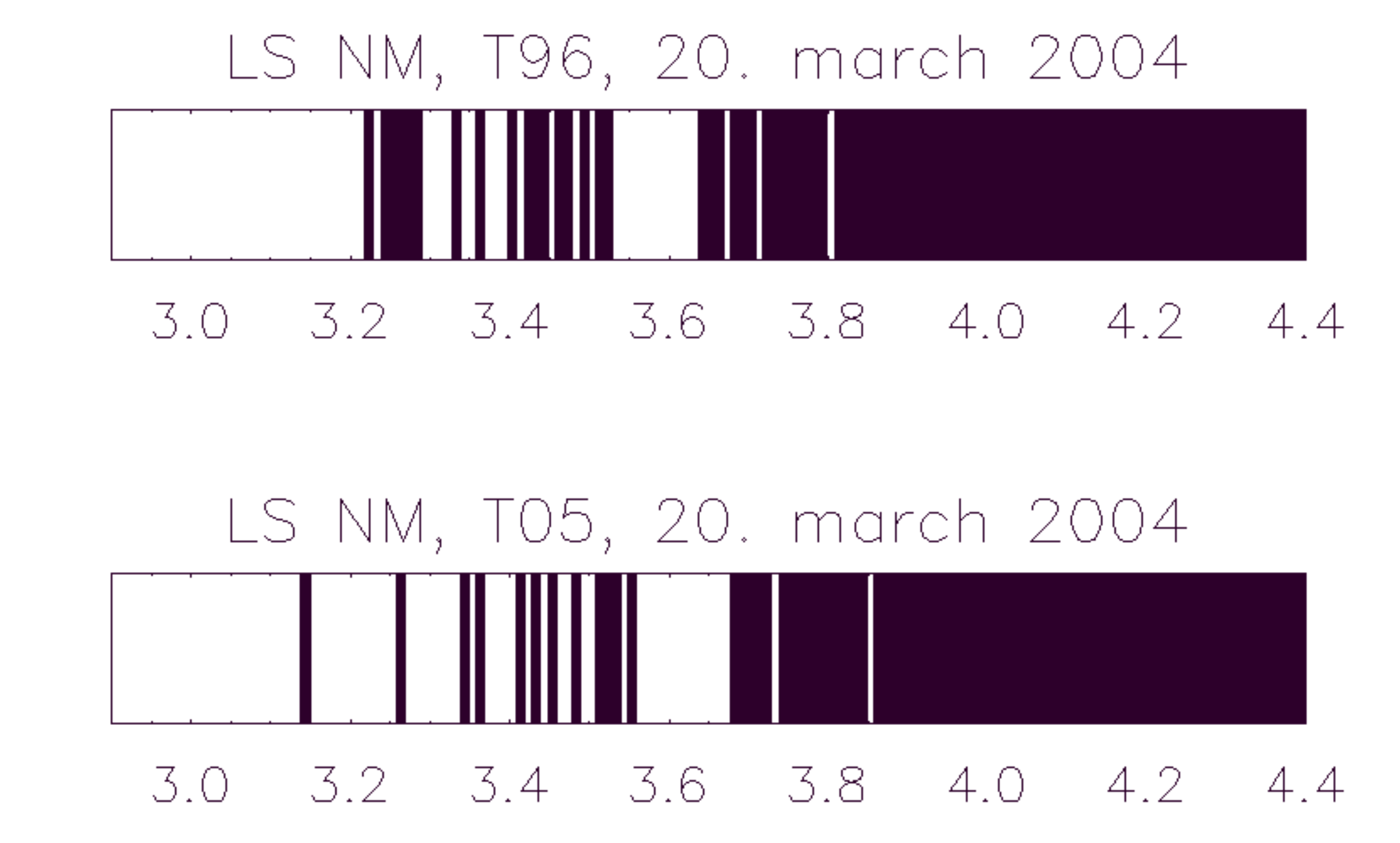}
  \caption{Spectrum of allowed rigidities for Lomnicky Stit neutron monitor}
  \label{simp_fig1a}
\end{figure}

Other examples of model results are on the Fig. \ref{simp_fig1a}, where spectrum of allowed rigidities for Lomnicky Stit neutron monitor station for 20. march 2004 
12:00 UTC 
in GeoMag96 (upper panel of figure) and GeoMag05 (bottom panel) is presented. Figure \ref{simp_fig1a} shows a typical structure of alowed trajectories spectrum for a midle latitude 
station with penumbra region and effective cut-off rigidity 3.5 GV in T96 external field and 3.58 GV in T05.

GeoMag code has been used both for reproducing known effects as East-West effect and rigidity cutoff calculations as well to 
separate Primary (allowed trajectories) from Secondary (forbidden ones) Cosmic Rays \cite{bib:ams_mib}. 

\subsection{Model precision}

GeoMag code can be used to estimate the CR intensity inside magnetosphere. The evaluation from vertically incoming particles, hovewer often used, 
is still an approximation, so the precision of this kind of approach should be considered. The precision of vertical approximation can be checked by comparison 
with simulations for all incoming directions. We realized a simulation for selected neutron monitors positions at the Earth surface namely 
Lomnicky Stit (49.20$^{\circ}$ N, 20.22$^{\circ}$ E), Newark (39.68$^{\circ}$ N, 75.75$^{\circ}$ W) and Oulu (65.05$^{\circ}$N, 25.47$^{\circ}$E) \cite{bib:nmdb} .
Instead of one vertical incoming particles, 576 incoming directions were back traced for every station with our code using Tsyganenko 96 external field. 
Incoming directions cover a half sphere, where for every direction we evaluate 20 thousand energies, starting from 0.01GV with 0.01 GV step. 
The intensity was evaluated from spectrum at 1 AU provided by HelMod model. 

\begin{figure}[ht!]
  \centering
   \includegraphics[width=0.45\textwidth]{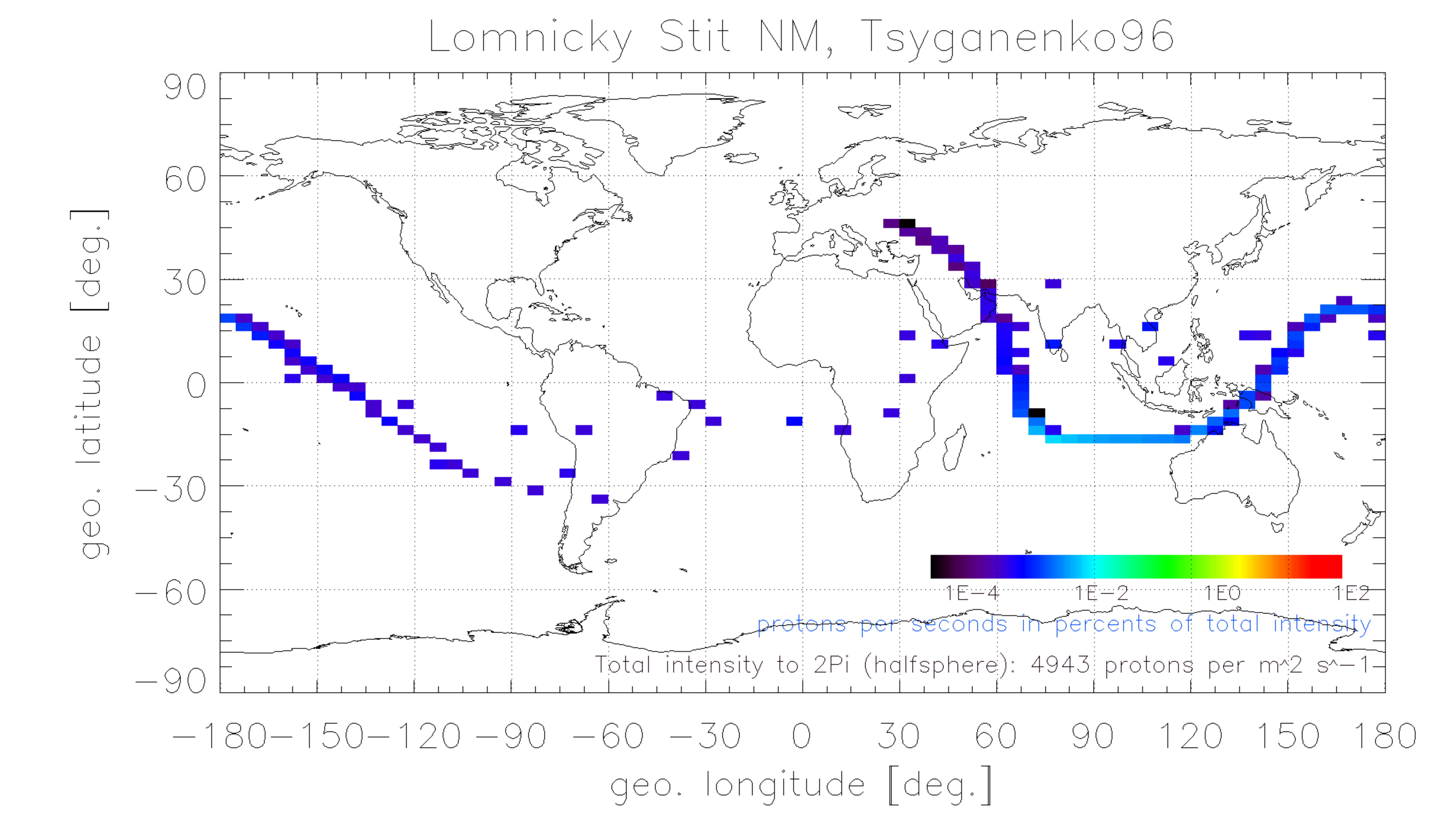}
   \includegraphics[width=0.45\textwidth]{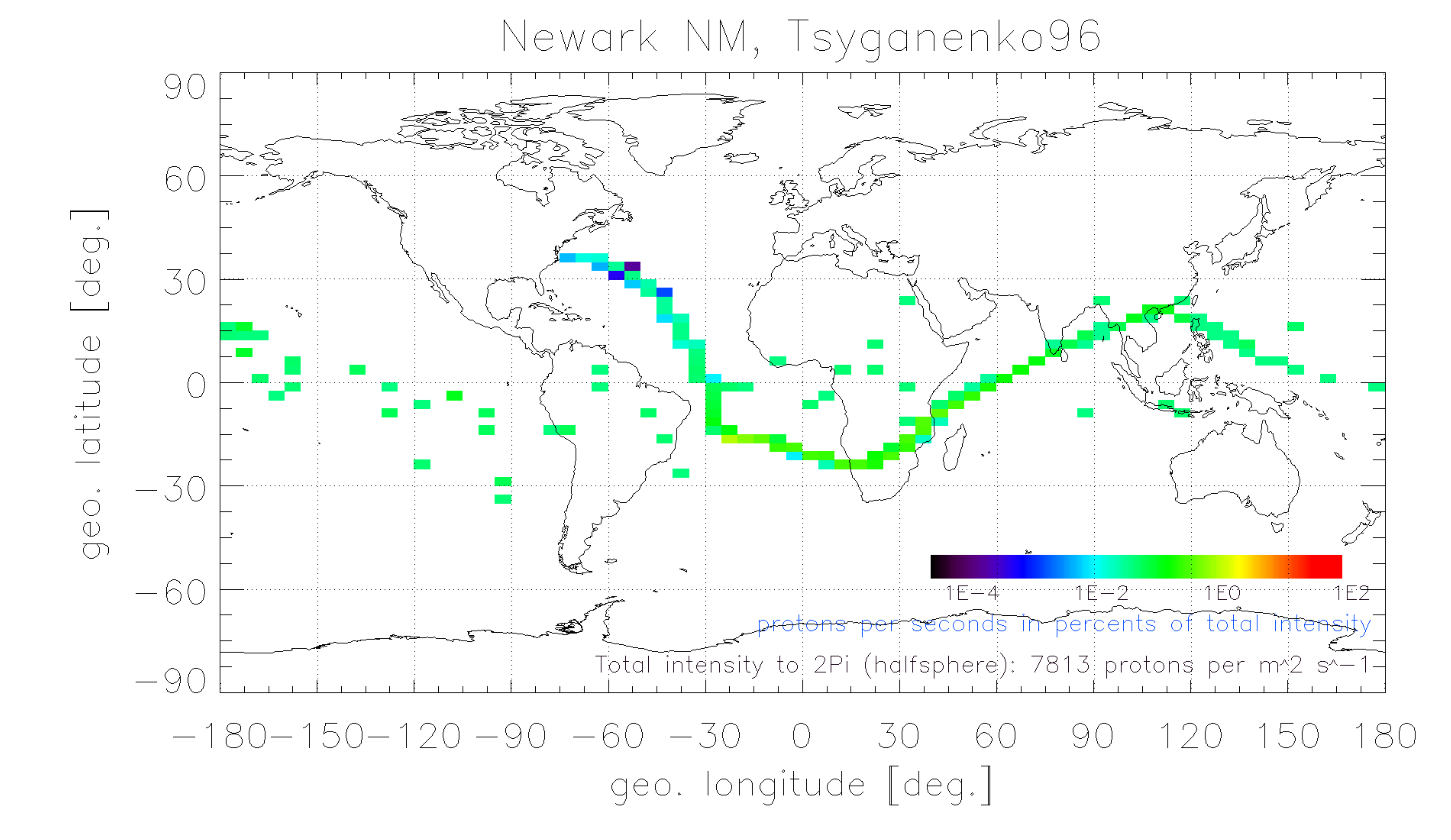}
   \includegraphics[width=0.45\textwidth]{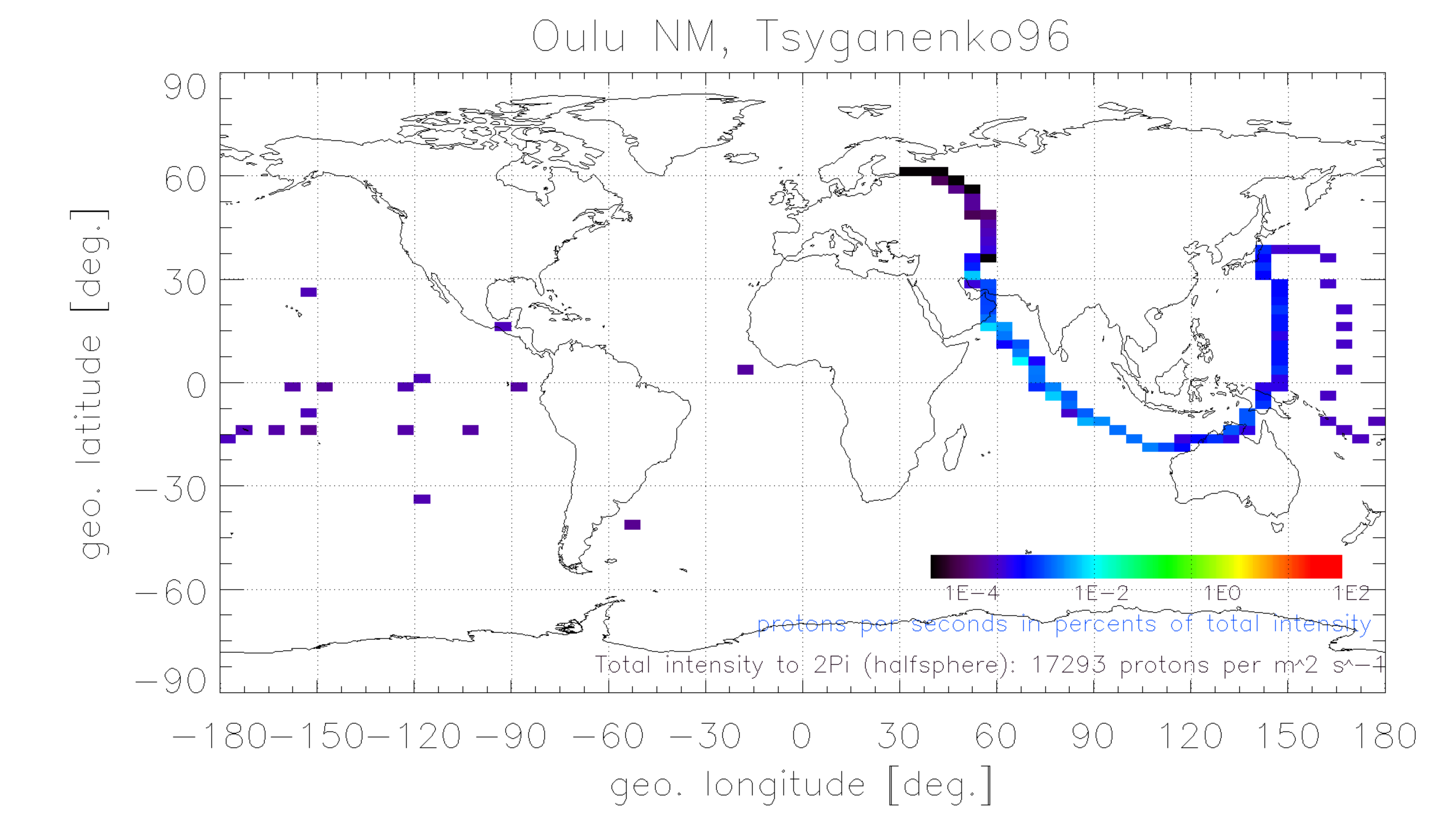}
  \caption{Proton intensities for NM stations Lomnicky stit, Newark and Oulu evaluated by vertical approach.}
  \label{simp_fig2}
 \end{figure}

\begin{figure}[ht!]
  \centering  
  \includegraphics[width=0.45\textwidth]{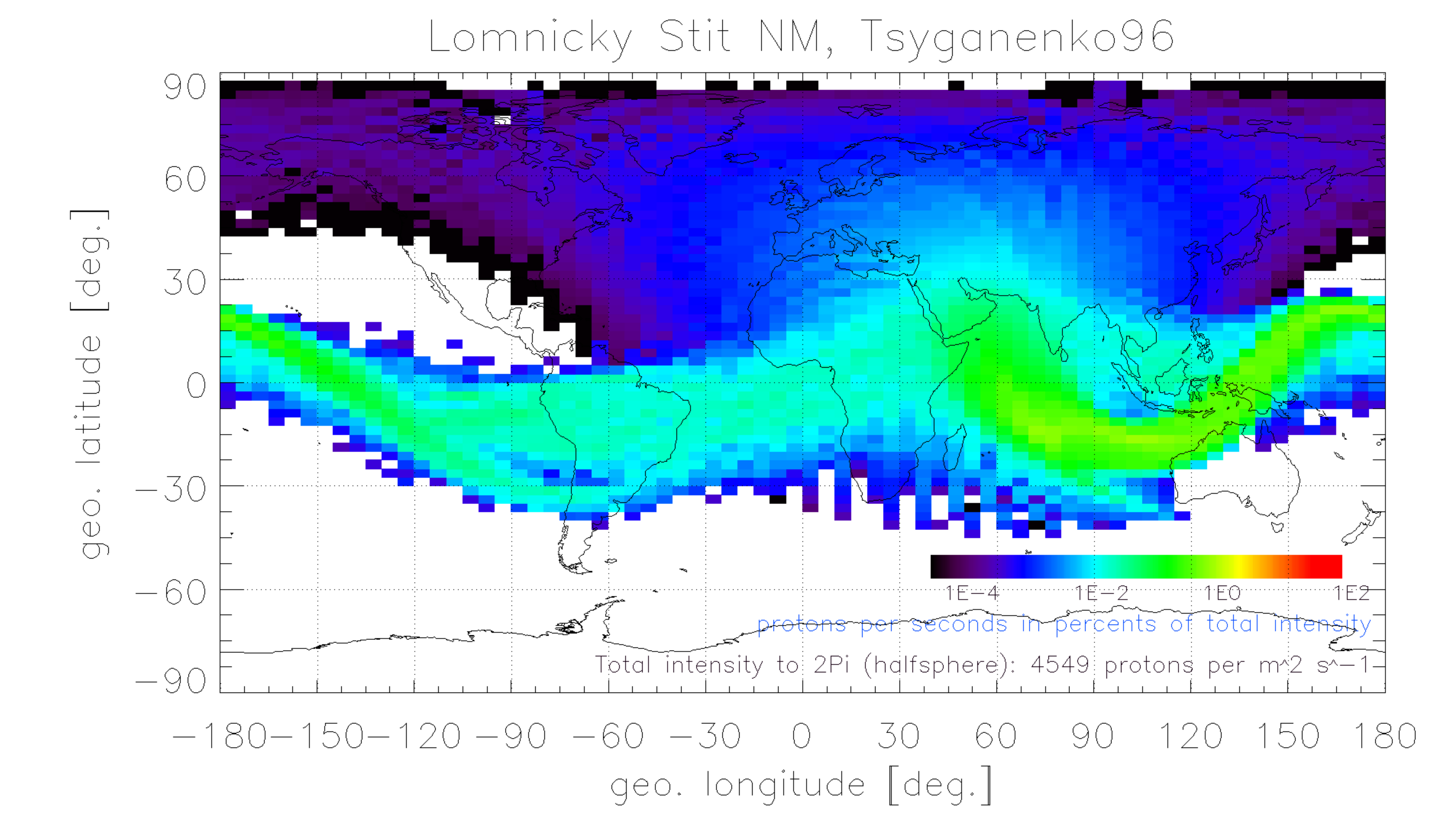}
  \includegraphics[width=0.45\textwidth]{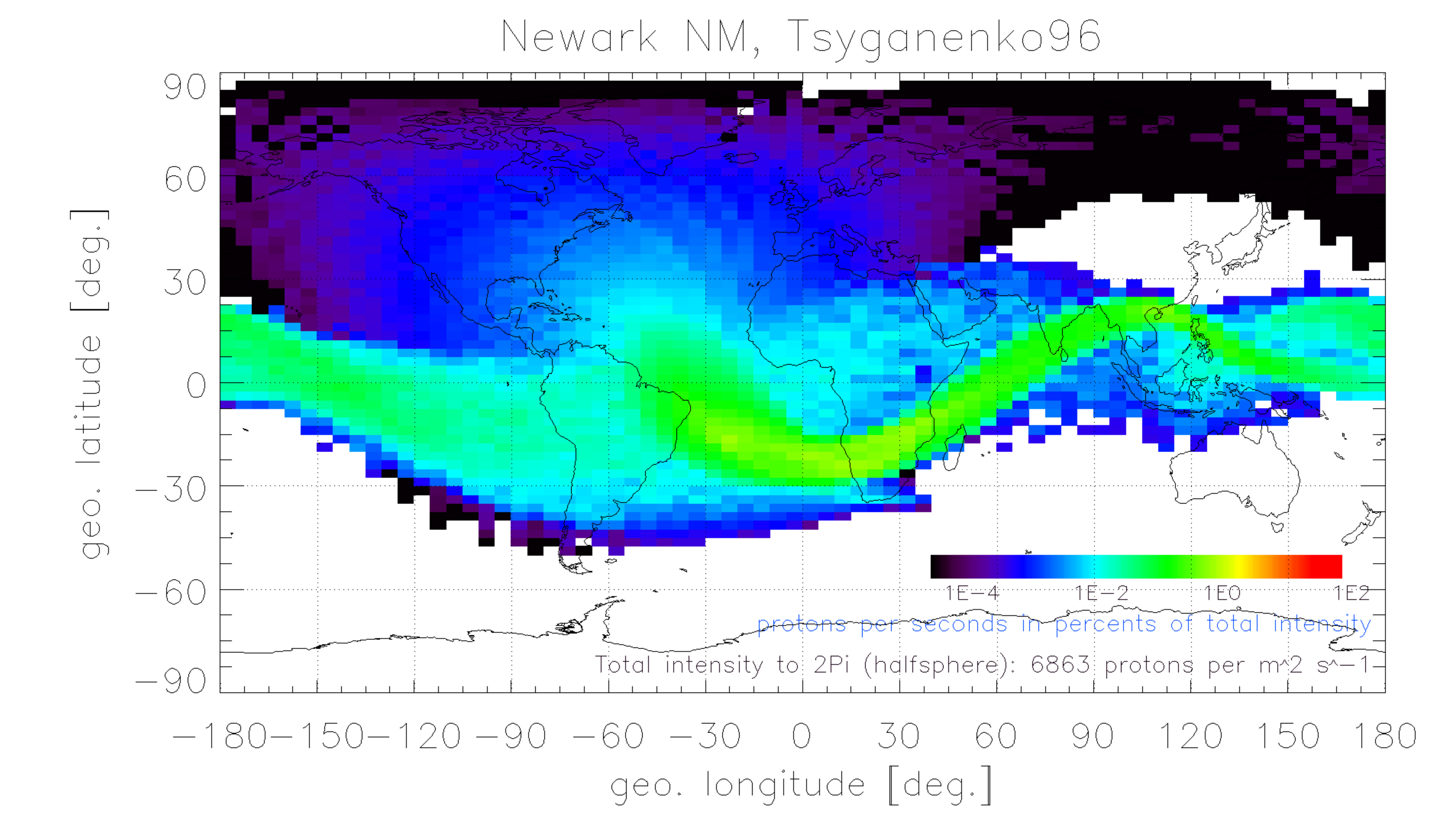}
  \includegraphics[width=0.45\textwidth]{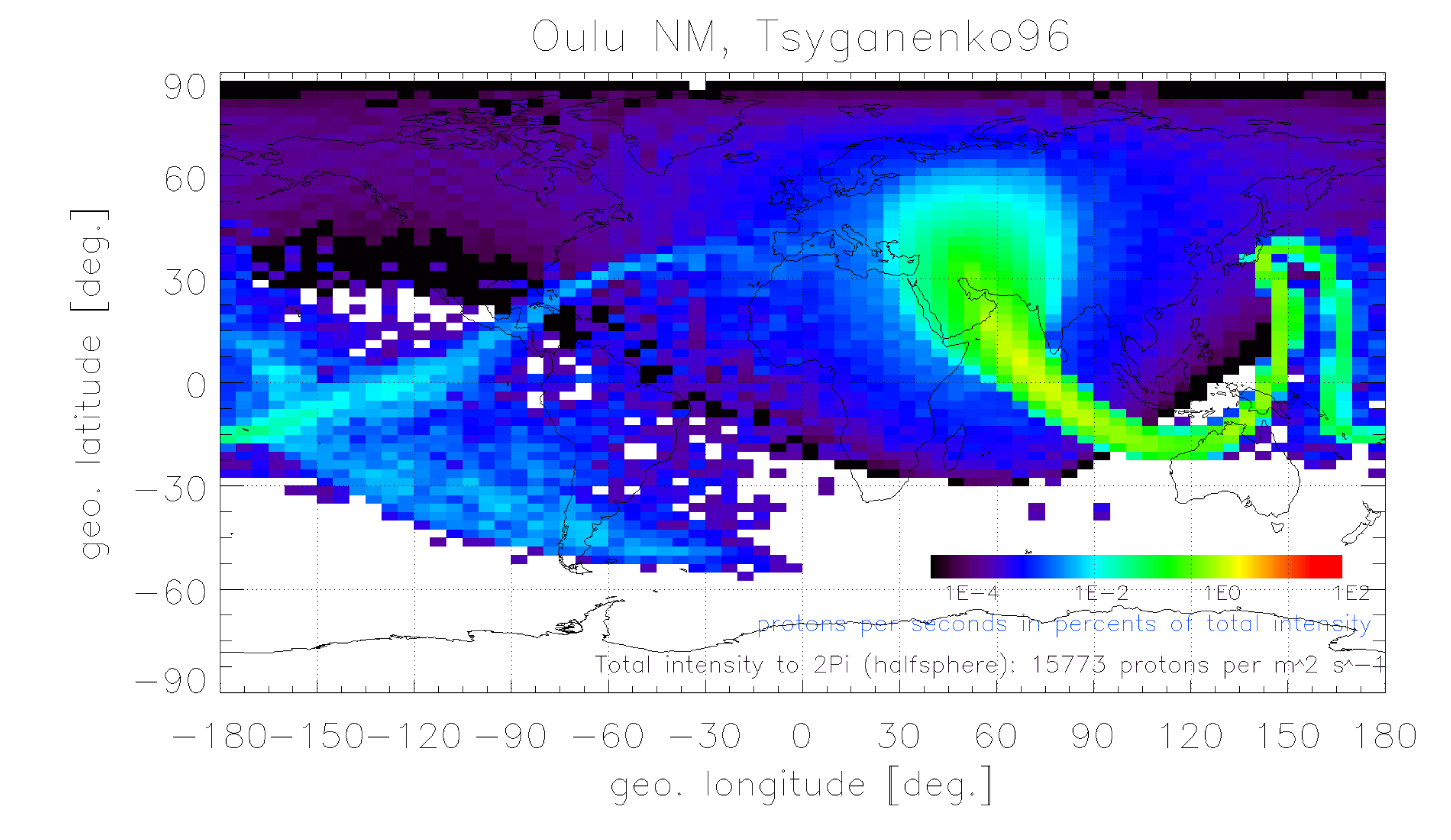}
  \caption{Proton intensities for NM stations Lomnicky stit, Newark and Oulu evaluated by multidirectional approach.}
  \label{simp_fig3}
 \end{figure}

In Figures \ref{simp_fig2} and \ref{simp_fig3} we present the difference 
between vertical approach and  all incoming directions approach. 

The all direction approach shows the full asymptotic cone of NM station.
Intensity of incoming particles from different asympotic directions is
indicated by colors representing a part of the total intensity from full
cone in \%.


For Lomnicky Stit station the difference in proton total intensity (to full sphere) evaluated by both methods is 8\%. 
For Newark station 12 \% and 9\% for Oulu. The evaluated intensities are presented in Table \ref{table_1}.

\begin{table}[ht!]
\begin{center}
\begin{tabular}{|c|c|c|c|}
\hline  & Lomnicky Stit & Newark & Oulu\\ \hline
All dir/T96  & 4549 & 6863 & 15773\\ \hline
Vertical dir/T96   & 4943 & 7813 & 17293\\ \hline
\end{tabular}
\caption{Intensities for different NM stations in $\frac{protons}{m^{2}s\cdot 2\pi}$}
\label{table_1}
\end{center}
\end{table}


\begin{figure}[ht!]
  \centering  
  \includegraphics[width=0.45\textwidth]{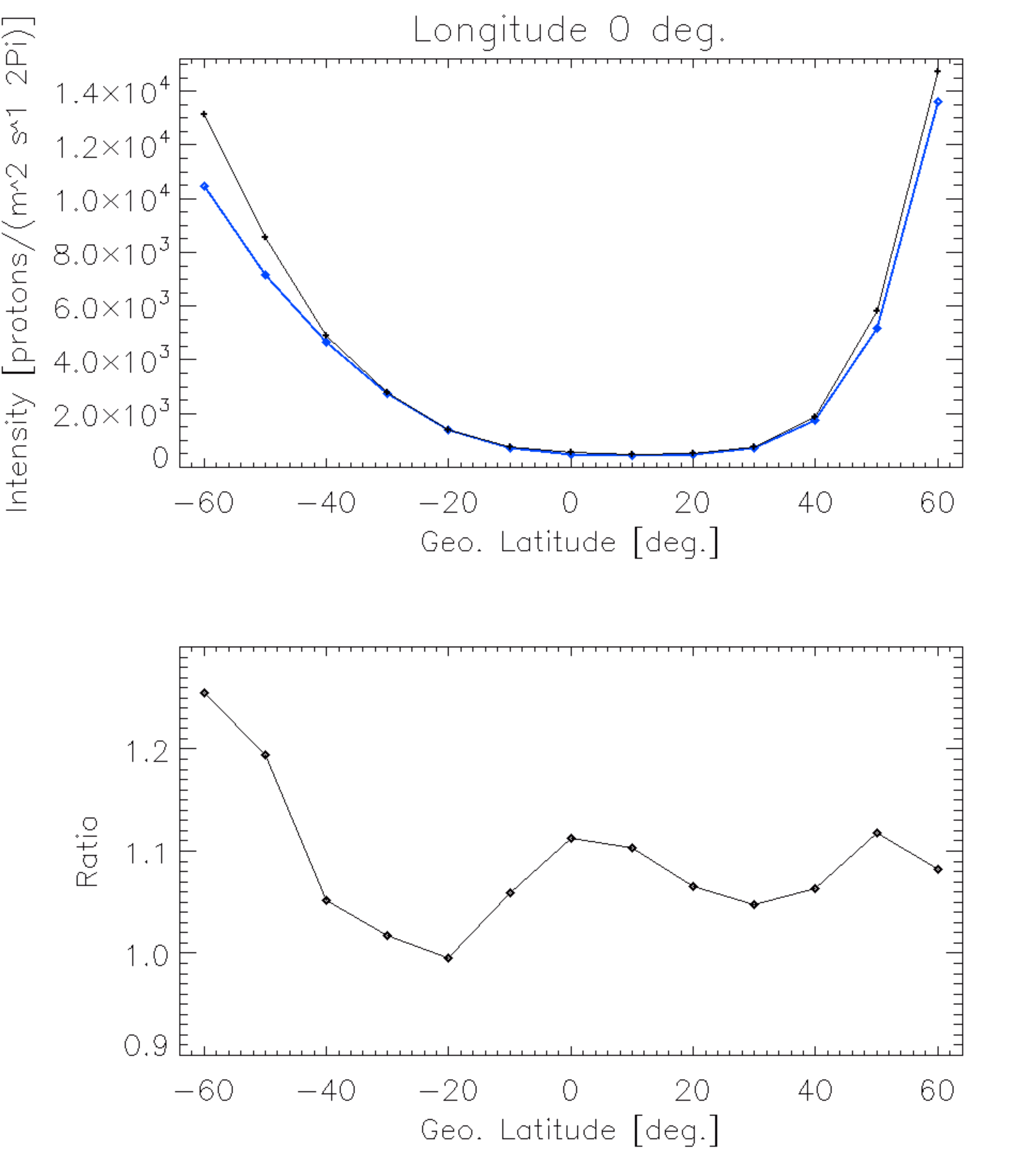}
  \caption{Proton intensities evaluated by vertical and multidirectional approach along meridional line with latitude 0 degrees (upper panel). 
           Ratio between intensities (bottom line, see text for description).}
  \label{simp_fig4}
 \end{figure}

For comparison in global scale we simulated a set of 13 points along a meridional line with longitude fixed at zero. Selected positions cover 
latitudes from 60 till -60 deg. with 10 deg. latitudinal step. Difference between intensity of particles evaluated by vertical 
approach and all directions approach for meridian line is showed in Fig. \ref{simp_fig4}. The blue line with diamonds on the Figure represents the 
intensities evaluated by all directions approach. Black line represent the intensities evaluated by vertical approach. 
Simulations show that while for low latitudes the vertical ap-
proach is relatively precise (around 10\% percent), for middle
and higher latitudes the difference can increase up to 25 \%.

Simulation of CR intensity from all directions take longer time than for one incoming direction 
(minutes vs. day of CPU time on 1 core).

\section{HelMod.org}

\begin{figure}[ht!]
  \centering
  \includegraphics[width=0.45\textwidth]{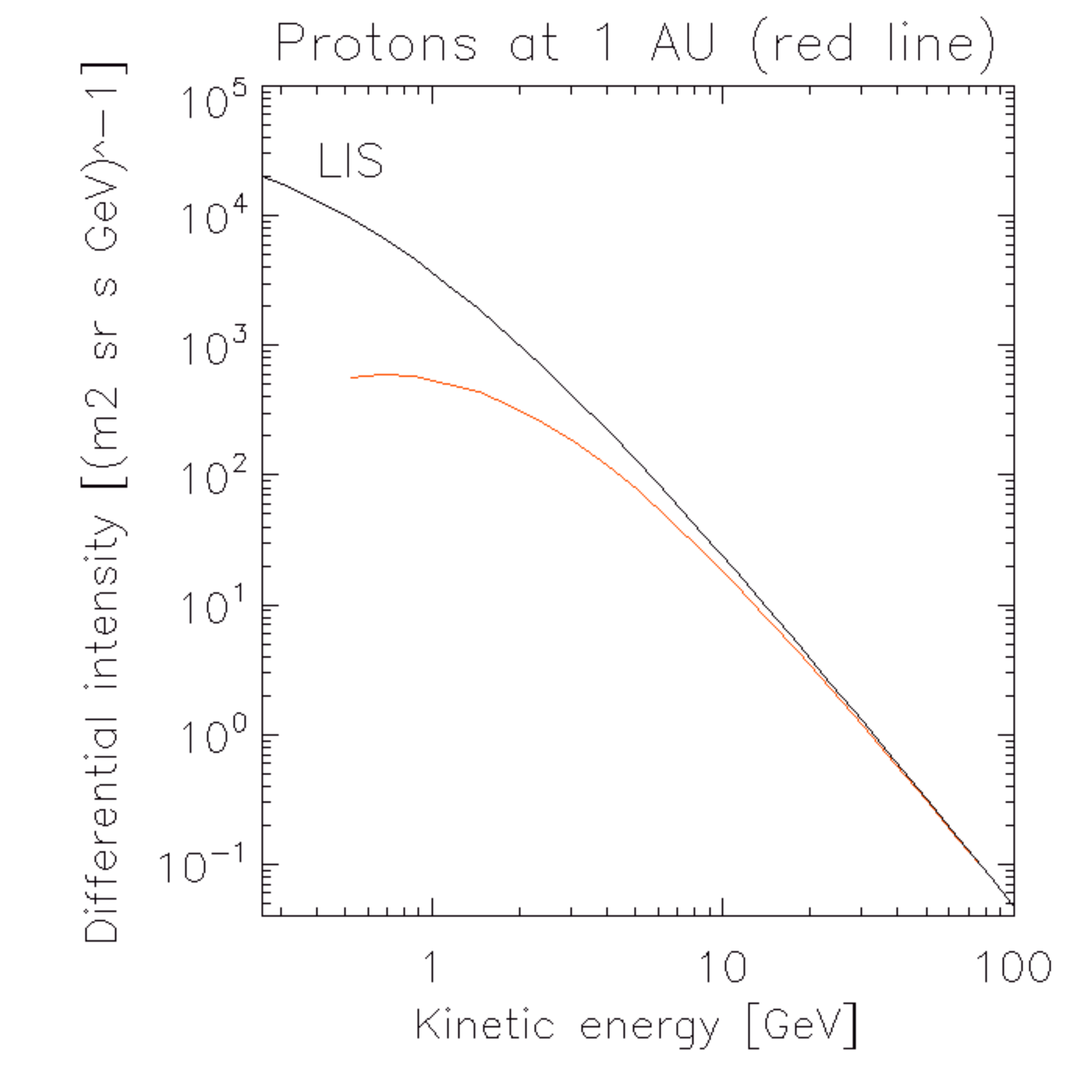}
  \includegraphics[width=0.45\textwidth]{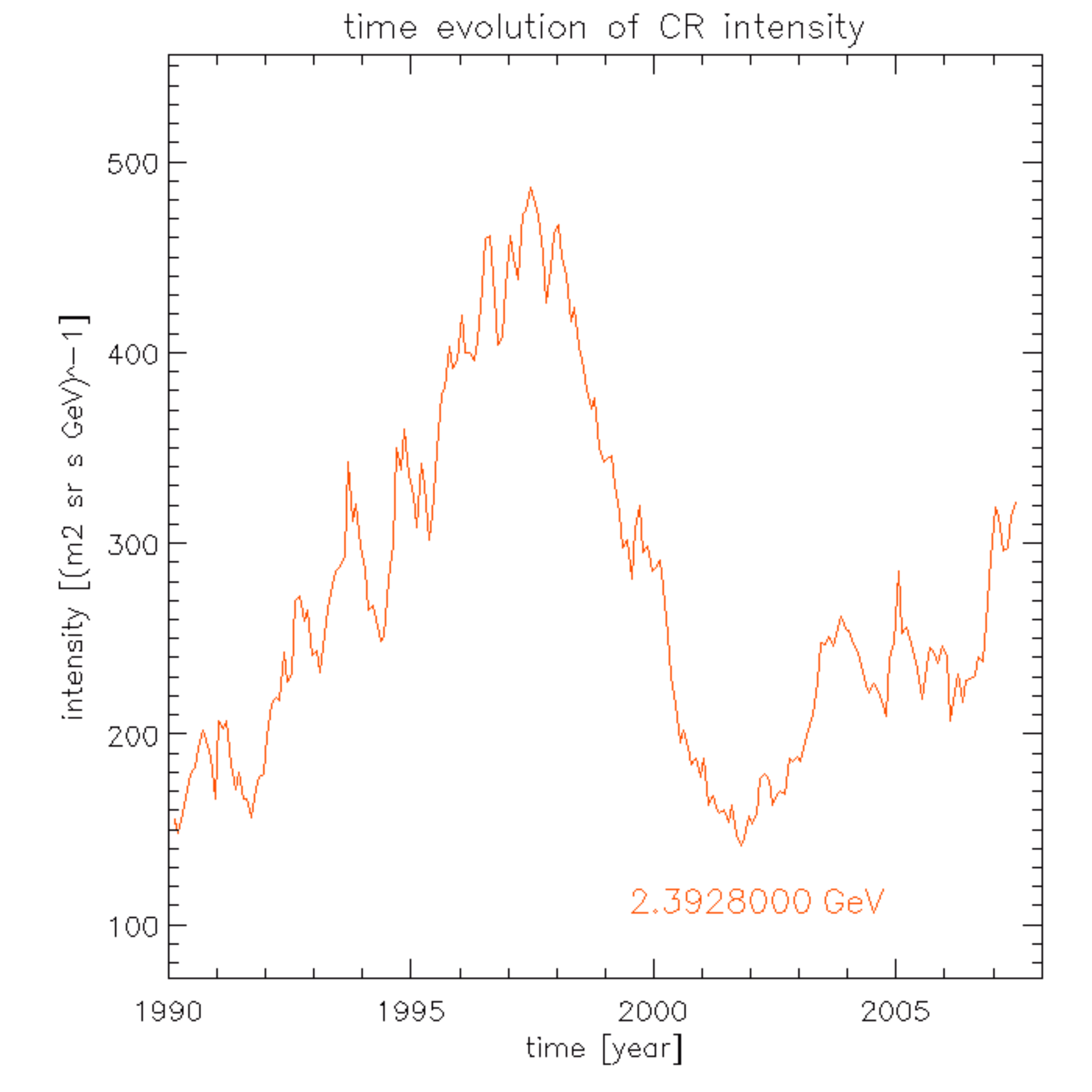}
  \caption{Protons differential spectrum from may 2005 (upper panel) at 1AU. Result of simulation in HelMod model. Evolution of 2.3 GeV protons (bottom panel) as evaluated by HelMod model.}
  \label{simp_fig5}
 \end{figure}

This Model uses a Monte Carlo Approach to solve the Parker Transport Equation, obtaining a modulated proton flux for a period 
(monthly averaged) between January 1990 and December 2007. 
Precise model description can be found in \cite{bib:mib_art}\cite{bib:art_icrc}.
The model HelMod (Heliospheric Modulation model) at helmod.org is a catalog of proton spectra at 1AU. Spectra were evaluated in 
Helmod version 1.5. for 18 consecutive years from 1990 to 2008, including the full 23$^{rd}$ solar cycle.
By setting a time in the form of month and year on the web interface, the end user can get differential spectrum of protons or time evolution of 
protons intensity in one of available energetic channels (from bin with center at 0.52 GeV to 177 GeV). 
Energetic channels are identical to AMS-01 mission channels \cite{bib:ams01}. 
Example of HelMod results are shown in Fig. \ref{simp_fig5}. In upper panel there is the proton spectrum at 1AU from may 2005, bottom panel shows 
the evolution of 
proton intensity with energy 2.4GeV in a period from 1990 till 2007.
 
\section{Conclusions}

We present the models GeoMag and HelMod web versions, describing galactic cosmic ray propagation through heliosphere and magnetosphere.
geomagsphere.org allows user to run a code for two models of geomagnetic field (Tsyganenko 96 and 05) and evaluate cut-off rigidity and set 
of characteristics for every allowed vertical trajectory for selected point in the magnetosphere. Precision of vertical approximation 
was evaluated for three selected neutron monitors sites.

helmod.org provide catalog of HelMod results (proton spectra) at 1AU for 18 years consecutive years since 1990 till 2007.

\section{Acknowledgements}
This work was supported by VEGA grant agency project 2/0076/13.
This work is supported by Agenzia Spaziale Italiana under contract ASI-INFN I/002/13/0, Progetto AMS - Missione scientifica ed analisi
dati.

\vspace*{0.5cm}

\end{document}